\documentclass[fleqn,usenatbib]{mnras}

\usepackage[utf8]{inputenc}
\usepackage{enumitem}
\usepackage{graphicx}
\usepackage{hyperref}
\usepackage{mathtools}
\usepackage{comment}
\usepackage{amsmath}
\usepackage{ amssymb }
\usepackage{lipsum} 
\usepackage{float}
\usepackage{cleveref}
\usepackage[caption=false]{subfig}
\usepackage{placeins}
\usepackage{xcolor}

\newcommand{\beq}{\begin{equation}}
\newcommand{\eeq}{\end{equation}}
\newcommand{\beqa}{\begin{eqnarray}}
\newcommand{\eeqa}{\end{eqnarray}}
\def\msun{\rm{M}_{\odot}}

\definecolor{notecolor}{rgb}{0.8,0,0}
\definecolor{OrangeRed}{HTML}{FF4500}
\definecolor{MidnightBlue}{HTML}{191970}
\definecolor{MyGreen}{rgb}{0.3,0.8,0.3}
\definecolor{YellowGreen}{HTML}{9ACD32}
\definecolor{Indigo}{HTML}{4B0082}

\makeatletter
\def\apptablenumbers{\global
\setcounter{table}{0}
\setcounter{figure}{0}
\setcounter{equation}{0}
\def\thetable{\thesection\the\c@table}%
\def\fnum@table{{\bf\tablename~\thetable}}%
\def\thefigure{\thesection\the\c@figure}%
}%

\title[Shocks in Cluster SZ Profiles]{Shocks in the Stacked Sunyaev-Zel'dovich Profiles of Clusters I: Analysis with the Three Hundred Simulations}

\author[E. J. Baxter et al.]{
Eric J.~Baxter,$^{1}$
Susmita Adhikari,$^{2,3,4}$
Jes\'us Vega-Ferrero,$^{5,6}$
Weiguang Cui,$^{7,8}$
\newauthor
Chihway Chang,$^{3,4}$
Bhuvnesh Jain,$^{9}$
and Alexander Knebe$^{10,11,12}$
\\$^{1}$Institute for Astronomy, University of Hawai'i, 2680 Woodlawn Drive, Honolulu, HI 96822, USA
\\$^{2}$Kavli Institute for Particle Astrophysics and Cosmology and Department of Physics, Stanford University, Stanford, CA 94305, USA
\\$^{3}$Department of Astronomy and Astrophysics, University of Chicago, Chicago, IL 60637, USA
\\$^{4}$Kavli Institute for Cosmological Physics, University of Chicago, Chicago, IL 60637, USA
\\$^{5}$IFCA, Instituto de F\'{\i}sica de Cantabria (UC-CSIC), Av. de Los Castros s/n, 39005 Santander, Spain
\\$^{6}$Artificial Intelligence Research Institute (IIIA), Campus UAB, 08193 Bellaterra, Spain
\\$^{7}$Departamento de F\'{\i}sicaa Te\'{o}rica, M\'{o}dulo 15, Facultad de Ciencias, Universidad Aut\'{o}noma de Madrid, E-28049 Madrid, Spain
\\$^{8}$Institute for Astronomy, University of Edinburgh, Royal Observatory, EH9 3HJ Edinburgh, United Kingdom
\\$^{9}$Department of Physics and Astronomy, Center for Particle Cosmology, University of Pennsylvania, Philadelphia, PA 19104, USA
\\$^{10}$Departamento de F\'isica Te\'{o}rica, M\'{o}dulo 15, Facultad de Ciencias, Universidad Aut\'{o}noma de Madrid, 28049 Madrid, Spain
\\$^{11}$Centro de Investigaci\'{o}n Avanzada en F\'isica Fundamental (CIAFF), Facultad de Ciencias, Universidad Aut\'{o}noma de Madrid, \\ 28049 Madrid, Spain
\\$^{12}$International Centre for Radio Astronomy Research, University of Western Australia, 35 Stirling Highway, \\ Crawley, Western Australia 6009, Australia
}

\begin{document}
\label{firstpage}

\pagerange{\pageref{firstpage}--\pageref{lastpage}}
\maketitle

\begin{abstract}
Gas infalling into the gravitational potential wells of massive galaxy clusters is expected to experience one or more shocks on its journey to becoming part of the intracluster medium (ICM).  These shocks are important for setting the thermodynamic properties of the ICM and can therefore impact cluster observables such as X-ray emission and the Sunyaev-Zel'dovich (SZ) effect.  We investigate the possibility of detecting signals from cluster shocks in the averaged thermal SZ profiles of galaxy clusters.  Using zoom-in hydrodynamic simulations of massive clusters from the Three Hundred Project, we show that if cluster SZ profiles are stacked as a function of $R/R_{200m}$, shock-induced features appear in the averaged SZ profile.  These features are not accounted for in standard fitting formulae for the SZ profiles of galaxy clusters.  We show that the shock features should be detectable with samples of clusters from ongoing and future SZ surveys.  We also demonstrate that the location of these features is correlated with the cluster accretion rate, as well as the location of the cluster splashback radius.  Analyses of ongoing and future surveys, such as SPT-3g, AdvACT, Simons Observatory and CMB-S4, that include gas shocks will gain a new handle on the properties and dynamics of the outskirts of massive halos, both in gas and in mass.
\end{abstract}


\section{Introduction}

In the standard picture of structure formation, gas falling into galaxy clusters experiences one or more shocks before becoming part of the intracluster medium (ICM).  The shocking process is essential to the conversion of the kinetic energy associated with infall into the thermal energy of the gas, and is therefore at least partly responsible for setting the thermodynamic properties of the ICM \citep{Evrard:1990}.  For this reason, shocks are closely connected to our interpretation of cosmological observables tied to the ICM, such as the X-ray and Sunyaev Zel'dovich (SZ) signals from clusters \citep[e.g.][]{Lau:2015}, and the SZ power spectrum.  The properties and geometry of cluster shocks are connected to the large-scale cluster environment \citep{Molnar:2009} and to processes like accretion and mergers \citep{Zhang:2020}. A better understanding of cluster shocks --- in both theory and data --- is therefore important both from a cosmological perspective and for understanding how clusters grow and evolve over time.  Additionally, cluster shocks are also thought to be sites of cosmic-ray acceleration \citep{Kang:1996}. 

Gas shocks near clusters can be divided into several (not necessarily mutually exclusive) classes depending on their physical origin and properties.  {\it Accretion} shocks can be defined as the shocks associated with the deceleration of baryons that are infalling into a cluster \citep{Bertschinger:1985}.  In contrast, {\it merger} shocks are generated by the infall of massive substructure merging with the main halo, which can result in an an outwardly propagating shock.  Recently, \citet{Zhang:2020} have pointed out that sufficiently strong merger shocks can propagate to large distances, forming so-called {\it merger-accelerated accretion} shocks.  Cluster shocks can also be divided into two classes based on their relative location: {\it external} and {\it internal} shocks.  External shocks are associated with pristine, previously unshocked gas, falling onto the cluster and being shocked for the first time \citep{Lau:2015}.  Internal shocks can result from continued infall of gas after an external shock, from mergers of multiple halos, and from bulk flows \citep{Molnar:2009}.   External accretion shocks typically occur at several virial radii and have high Mach number (roughly 50-300), while internal accretion shocks are typically found near the virial radius and have low Mach number (less than about five) \citep[e.g.][]{Molnar:2009}.   Internal accretion shocks are sometimes referred to as {\it virial} shocks.  We note that these categorizations of shocks are complicated by the fact that shocks often appear very differently in simplified spherically symmetric simulations relative to realistic cosmological simulations in which the shock geometry is impacted by, for instance, the presence of filaments.  Indeed, in spherically symmetric simulations, gas experiences a single accretion shock near the splashback radius \citep{Bertschinger:1985, Shi:2016}.

Here we do not attempt to distinguish between the different physical origins of shocks, but rather study their impact on the SZ profiles of galaxy clusters.  We focus mainly on external shocks at several virial radii that likely result from merger-accelerated accretion shocks.  Recent studies have shown that such shocks are nearly ubiquitous in the outskirts of galaxy clusters \citep[e.g.][]{Zhang:2020,Aung:2020}.  For simplicity, we will refer to these as external shocks.\footnote{We note that some authors refer to such external shocks as accretion shocks \citep[e.g.][]{Molnar:2009}.}  We will also occasionally consider internal accretion shocks, i.e. virial shocks.

In spherically symmetric, self-similar models of halo formation, the location of the external shock is related to the current turnaround radius of the cluster (i.e. the radius at which matter stops expanding with the Hubble flow and begins infall towards the cluster) as well as to the halo accretion rate \citep{Bertschinger:1985, Shi:2016}.  Self-similar models, however, are not expected to describe real clusters for a few reasons.  First, the effects of dark energy at late times induce departures from this self-similar picture  \citep{Lau:2015}.  Moreover, in  realistic hydrodynamic simulations, where halos deviate from spherical symmetry and accretion is not smooth, the shock geometry can be complicated, with shocks having different radii depending on direction \citep[e.g.][]{Evrard:1990,Skillman:2008,Molnar:2009,Power2020}.  

While external shocks have long been predicted by theory and simulations, detection of these shocks around galaxy clusters in data is complicated by the low density of gas in the cluster outskirts.  X-ray observations of thermal bremsstrahlung emission from clusters, for instance, scales as the square of the gas density, making its detection in cluster outskirts challenging \citep{Reiprich:2013}.  X-ray and $\gamma$-ray emission is also expected from cluster shocks as a result of electrons being highly accelerated in the shock fronts and inverse Compton scattering with cosmic microwave background (CMB) photons \citep[e.g.][]{Loeb:2000}.  Detection in $\gamma$-rays is challenging because of the small number of photons produced by the shocks and the low resolution of $\gamma$-ray telescopes  \citep{Keshet:2003}.  There have been several reported detections of gamma-ray rings associated with cluster virial shocks around individual clusters \citep[e.g.][]{Keshet:2017, Keshet:2018}, as well as in stacked gamma-ray observations \citep{Reiss:2018}.  The electrons accelerated in shock fronts can also produce a synchrotron signal in the radio \citep{BRUNETTI:2014}.  

A promising avenue for detecting cluster shocks is via the thermal Sunyaev-Zel'dovich effect \citep[SZ; ][]{SZ}.  The SZ effect is caused by inverse Compton scattering of CMB photons with hot electrons, leading to a spectral distortion of the CMB.  The amplitude of the SZ spectral distortion, typically expressed in terms of the Compton $y$ parameter, is proportional to a line of sight integral of the electron gas pressure.  Unlike X-ray observations --- which are sensitive to the square of the gas density --- the tSZ scales linearly in the gas density, making it a powerful probe of gas in the cluster outskirts where shocks are expected \citep[e.g.][]{Kocsis:2005}. \citet{Kocsis:2005} and \citet{Molnar:2009} have shown that the shock radius in simulated clusters is associated with a rapid change in the pressure profiles of these objects, which in turn leads to an SZ signal that could be identified in high-resolution observations from e.g. ALMA.\footnote{\url{https://www.almaobservatory.org/}}  

Several attempts have been made at detecting the SZ signatures from cluster shocks in data.  \citet{Basu:2016} report a detection of a cluster merger shock using high resolution observations with ALMA.  Similar measurements have been made by \citet{Ueda:2018} and \citet{DiMascolo:2019}.  \citet{Hurier:2017} claim detection of a virial shock around Abell 2319 using much lower resolution, and lower sensitivity {\it  Planck} data. \citet{Keshet:2020} report that measurements of virial shocks in the SZ are coincident with signals found in gamma-rays.

In this work, we investigate the observational prospects for identifying shock features in SZ observations around large cluster samples from current and future wide-field SZ cluster surveys. Ongoing and planned CMB surveys --- including SPT-3G \citep{Benson:2014}, AdvACT \citep{Henderson:2016}, Simons Observatory \citep{Ade:2019}, and CMB Stage 4 (CMB-S4; \citealt{S4}) --- will map large fractions of the submillimeter sky to sensitivities 10 to 30 times higher than the {\it Planck} satellite \citep{Planck:y}, and with better angular resolution.  Furthermore, unlike ultra-high resolution observations e.g. ALMA, these wide-field surveys will naturally obtain measurements to very large cluster-centric radii with their wide fields of view.  Given that these surveys are expected to detect many tens of thousands of galaxy clusters \citep{S4} it is useful to consider whether we can detect shock features around these clusters, and what we can learn from such measurements.

As we show below, we are unlikely to obtain high significance measurements of external shocks around {\it individual} clusters with current and near-term wide-field SZ survey data.  However, by combining measurements from many clusters, it may be possible to enhance the signal-to-noise.  We show that although there is significant scatter in shock positions and amplitude from cluster to cluster, the $y$ profile averaged over many clusters does contain distinct features connected to external shocks, and that these features are detectable with near-term surveys.   

External shocks form in some sense the boundary of the gaseous component of the halo, separating infalling gas from the higher temperature gas inside shock boundary. Analogous to the gas boundary, the dark matter component has a boundary at the splashback radius, i.e the surface formed by the apocenters of the most recently accreted particles \citep{Diemer:2014xya, AD14, More:2015ufa, Shi:2016}.   In an idealized spherical solution there is a density caustic at this location \citep{1984ApJ...281....1F, Bertschinger:1985}.  Self-similar collapse solutions predict that the location of the external shock and the dark matter caustic coincide for moderately accreting halos when the gas has an adiabatic index of $\gamma = 5/3$, with the shock radius decreasing with accretion rate and increasing with adiabatic index \citep{Shi:2016}.  Realistic hydrodynamical simulations, however, find that the shock can extend much further outside the splashback boundary of a halo \citep[e.g.][]{Lau:2015}. As both the shock radius and the splashback radius correlate strongly with the accretion rate of the halo, self-similar models and hydrodynamical simulations both predict that the cluster shock and splashback radii are correlated \citep{Bertschinger:1985,Lau:2015,Shi:2016, Walker:2019, Aung:2020}.  \citet{Hurier:2017} use a comparison between the measured shock and splashback radii of a massive, low-redshift cluster detected by {\it Planck} to derive a constraint on the gas adiabatic index.  The boundary of the gas component and its relation to the dark matter boundary is relevant to understanding galaxy evolution in clusters, as this boundary delineates the onset of intra-cluster processes.    We investigate correlations between shock features in the cluster-averaged SZ profiles and the cluster splashback radius.  This correlation is potentially measurable in data given a suitable proxy for the cluster accretion rate.

Our analysis relies on a suite of hydrodynamical simulations from the Three Hundred Project\footnote{\url{https://the300-project.org/}}.  These are a set of re-simulations of 324 cluster-centered regions extracted from the \textit{The MultiDark Planck 2} simulation\footnote{\url{https://www.cosmosim.org}} (hereafter MDPL2).  The Three Hundred project provides a large number of massive clusters, while also providing high resolution gas physics necessary to capture shocks. 

The paper is organized as follows.  In  \S\ref{sec:simulated_data} we describe the Three Hundred project simulations, how we extract Compton $y$ maps from the simulations, and how we use these to construct mock-observed Compton $y$ profiles.  We present our analysis of these simulated profiles in \S\ref{sec:results} and conclude in \S\ref{sec:discussion}.

\section{Simulated Data}
\label{sec:simulated_data}

\subsection{The Three Hundred Project}

The simulated galaxy clusters studied in this work belong to the Three Hundred Project, which is described in \cite{Cui2018}. The sample consists 324 spherical regions centered on each of the most massive clusters identified at $z=0$ within the \textsc{MDPL2} simulation \citep{Klypin2016}, a $(1\,h^{-1}\,{\rm Gpc})^3$ dark matter (DM) only simulation containing $3840^3$ DM particles and using {\it Planck} 2016 cosmological parameters \cite{Planck2016} ($\Omega_M=0.307$, $\Omega_b=0.048$, $\Omega_{\Lambda}=0.693$, $\sigma_8=0.823$, $n_s=0.96$, $h=0.678$). The clusters are initially selected according to their halo virial mass\footnote{In this paper, we indicate with $R_{\Delta_{c/m}}$ the radius of the sphere whose density is $\Delta_{c/m}$ times the critical density (denoted by $c$) or the mean density (denoted by $m$) of the Universe at that redshift $\rho(R_{\Delta_{c/m}})=\Delta\rho_{c/m}(z)$. We specifically use overdensities equal to $\Delta=500$,$200$, and vir, where $\Delta_{vir}$ corresponds roughly to 98 for the assumed cosmological model\citep{Bryan1998}. } $M_{\rm vir, c}\gtrapprox 8\times10^{14}h^{-1}M_\odot$ at $z=0$. However, a much larger region (beyond the Lagrangian volume of the cluster) with radius $R=15h^{-1}{\rm Mpc}$ and centered on the minimum of the cluster potential at $z=0$ is used to generate the initial conditions for each cluster with multiple levels of mass refinement using the \textsc{ginnungagap} code,\footnote{\url{https://github.com/ginnungagapgroup/ginnungagap}} i.e. the particle number at the outer region is decreased based on their distance to the cluster center to provide the gravitational field at the largest scales. These initial conditions are re-simulated with several simulation codes which use different baryon models and hydrodynamic solvers: \textsc{gadget-x} \cite{Murante2010,Rasia2015}, \textsc{gadget-music} \cite{MUSICI}, and \textsc{gizmo-simba} (Cui et al. 2020 in preparation). Given the large volume of the re-simulated regions (which exclusively include the cluster environment) and the multiple models employed to simulated the baryonic physics, these clusters have been used in several studies \citep[see ][]{Wang2018,Mostoghiu2019,Arthur2019,Ansarifard2020,Haggar2020,Kuchner:2020,Li:2020,Knebe:2020,Capalbo:2020,Vega-Ferrero:2020}.

The analysis in this work is based on the \textsc{gadget-x} model. It is a modified version of \textsc{gadget3} Tree-PM code and includes an improved SPH scheme with Wendland interpolating C4 kernel, and time dependent artificial thermal diffusion and viscosity. Some other main features of this simulation are the gas cooling with metal contributions, star formation with chemical enrichment with AGB phase \citep{Tornatore2007}, Supernovae and AGN feedback \citep{Steinborn2015}. The re-simulated clusters have a mass resolution  (in the high-resolution region) of $m_{\rm{DM}} = 1.27 \times10^9 h^{-1} \msun$ for the DM particles and $m_{\rm{SPH}} = 2.36 \times 10^8 h^{-1} \msun$ for the gas particles.  We note that the stability of simulated cluster entropy profiles to the choice of numerical code has been studied by \citet{Sembolini:2016}.

\subsection{$y$-maps from the Three Hundred simulations}

Compton $y$ maps for each cluster are generated with the \textsc{pyMSZ} package\footnote{\url{https://github.com/weiguangcui/pymsz}}, which has been presented in \cite{Cui2018} (see \citet{Baldi2018} for the kinetic SZ effect maps generated by this package). To generate the $y$-maps, we use all the simulation data within a radius of $12\,{\rm Mpc}/h$ from the center of the re-simulation region.  The signal is smoothed into a 2D-mesh with size of $5\,{\rm kpc}$ using the same SPH kernel as the simulation. This pixel size corresponds to an angular resolution of 2.5316", significantly higher resolution than the current and planned datasets that we consider.   In this work, we mainly focus on Compton $y$ maps generated from cluster snapshots a redshift of $z=0.193$.  This choice is motivated by the fact that ongoing and future SZ surveys will detect many clusters near this redshift, and it is sufficiently low that narrow shock features are likely to be resolved by roughly 1 arcminute resolution experiments.

We note that the simulations used to generate the $y$ maps do not treat the electron gas and ion gas separately, but rather consider both as a gas at a single temperature.  When the gas density is high, the electrons and ions reach thermal equilibrium quickly, making this a good approximation.  However, in the cluster outskirts relevant to this study, the gas density is low and the equilibration time can be large.  Since the ions carry most of the kinetic energy of infall due to their larger mass, shock heating results in the ions reaching significantly higher temperatures than the electrons and maintaining this temperature difference for a significant period of time \citep{Rudd:2009}.  Since it is the electrons that are primarily responsible for the SZ effect, the different temperatures of the ions and electrons can in principle impact the observables considered in this work.  \citet{Rudd:2009} used simulations to investigate the impact of this effect on SZ observations, finding for relaxed clusters a roughly 10\% effect on the $y$ profile near the shock radius.  Since we focus on relaxed clusters, and since the errorbars on the $y$ profiles in our analysis are significantly larger than 10\% near the shock radius (see, e.g., Fig.~\ref{fig:stacked_profile}), it is unlikely that this effect will significantly impact our main results.  We postpone a more careful investigation of this effect to future work.

\subsection{Cluster selection: identifying relaxed and not relaxed clusters}
\label{sec:relaxed}

As we show below, relaxed clusters can provide a cleaner environment in which to look for shock features in the cluster $y$ profiles than not relaxed clusters.  Recent mergers and significant substructure introduce features in the $y$ profiles that could interfere with our ability to identify features caused by shocks.  For the most part, therefore, we will focus our analysis on a subset of clusters deemed to be relaxed.  Since one of the main aims of this analysis is to determine the feasibility of detecting shock features in future data, we will adopt a relaxation criterion that can (at least approximately) be determined from data.  

We use the $f_{\rm sub}$ criterion (defined as the mass fraction in substructures identified within $R_{200c}$ of the main halo) to determine the relaxation state of the clusters, as this quantity can be inferred from observations through (satellite) galaxy-halo relations. The halo and galaxy catalogue was identified with the AMIGA Halo Finder \citep{AHF}. The lowest mass structure has at least 32 particles (including dark matter, gas and stars).
Our ``relaxed'' selection is 
\begin{equation}
    f_{\rm sub} < 0.1,
\end{equation}
which contains 68 clusters. We note here that $f_{\rm sub}$ correlates with the center of mass offset \citep{Cui2017}. Including center of mass offset as an additional parameter to quantify the cluster dynamical state only results in a small change to the sample.  As we discuss in more detail in \S\ref{sec:splashback}, $f_{\rm sub}$ correlates with the shock radius, as expected since higher $f_{\rm sub}$ halos will generally be faster accreters. The measured shock radius is therefore somewhat sensitive to our $f_{\rm sub}$ selection.  We also note that when stacking clusters, we throw out an additional six clusters for which $R_{200m}$\footnote{Note that because the matter density satisfies $\rho_m = \Omega_m \rho_{c}$ at $z=0$, $R_{200m}$ is much larger than $R_{200c}$.} is so large that we cannot reliably estimate the $y$ profile at large $R/R_{200m}$ because of the restrictions imposed by the finite high-resolution regions of the simulation. The mean mass of the removed clusters is $\langle M_{200m} \rangle = 1.27\times10^{15} M_{\odot}$.

\subsection{Mock observations}
\label{sec:mock_obs}

Several observational effects are expected to degrade our ability to detect shock features around galaxy clusters with future SZ data sets.  These include instrumental and foreground noise, the washing out of small-scale features by the telescope beam, and scatter in the mass-observable relations (which interferes with our ability to correctly scale the sizes of clusters when averaging their profiles).  We will explore the consequences of these effects by adding them to simulated data.

We incorporate $y$ noise into the analysis by generating Gaussian noise realizations of the expected instrumental and foreground noise for observations by CMB-S4.   The $y$ noise forecasts for CMB-S4 are taken from \citet{CMBS4_DSR}.

The impact of the telescope beam is simulated by convolving the two-dimensional Compton $y$ images with a Gaussian beam.  We assume $\sigma_{\rm beam} = 1\,{\rm arcmin}$, corresponding roughly to the resolution of $y$ maps that will be produced by CMB-S4.  As noted previously, all simulated clusters are at $z=0.193$.  Real clusters will of course be spread over some distribution in redshift, and will therefore experience different levels of beam smoothing in terms of physical scale.  However, as long as the cluster sample is at $z \lesssim 1$, we do not expect a significant impact of the beam on the identification of shock features, as we show below.  

When computing the averaged Compton $y$ profiles for the simulated clusters, we re-scale each profile in the radial direction by the value of $R_{200m}$ for each cluster (see discussion and motivation for using $R_{200m}$ in \S\ref{sec:profiles}).  This scaling ensures that the shock features occur in roughly the same location for each cluster, and significantly enhances the signal-to-noise of the shock features in the average $y$ profile.  However, for real data, the true cluster $R_{200m}$ is unknown.  Instead, we envision that a noisy estimate of the cluster mass will be formed from some SZ observable (such as the amplitude of $y$ within some aperture), and this estimate of cluster mass will be used to form an estimate of $R_{200m}$. We model the uncertainty in the cluster mass estimates by assigning each cluster an ``observed'' mass drawn from a Gaussian centered on the true cluster mass, $M_{200m}$ and with scatter $\sigma = 0.05 M_{200m}$, fairly typical of the level of scatter expected for the SZ observable $Y_{\rm SZ}$ \citep{Green:2020}.  These observed masses can then be used when estimating $R_{200m}$ for the purposes of stacking. 

\subsection{Compton $y$ profiles}
\label{sec:profiles}

We compute the azimuthally averaged Compton $y$ profiles of all the simulated clusters.  The profiles are first computed using 50 logarithmically spaced radial bins between 0.05 Mpc and 12 Mpc.  These profiles are then interpolated onto logarithmically spaced bins of  $R/R_{200m}$.  As discussed in \citet{Lau:2015}, the infall of gas onto a galaxy cluster after departure from the Hubble flow is determined by the enclosed mass at that time, which is set by the matter density.  Since the outer gas profiles of low redshift galaxy clusters are formed at late times during dark energy domination, the critical density of the Universe departs significantly from the matter density, and $\rho_m$ provides a better measure of the enclosed mass than $\rho_c$.  Therefore, the outer profiles of clusters are more self-similar when scaled by $R_{200m}$ rather than $R_{200c}$.  For the cases where we include observational scatter in the mass-observable relation, we use the mock-observed mass discussed above to compute $R_{200m}$.  Within each radial bin, we compute the average Compton $y$ value for the mock clusters to construct an azimuthally averaged $y$ profile.  The cluster $y$ profiles are then averaged across various subsets of clusters, as we discuss below.  The covariance of the average profile is computed using a leave-one-out jackknife. 

\begin{figure*}
    \centering
     \includegraphics[scale=0.6]{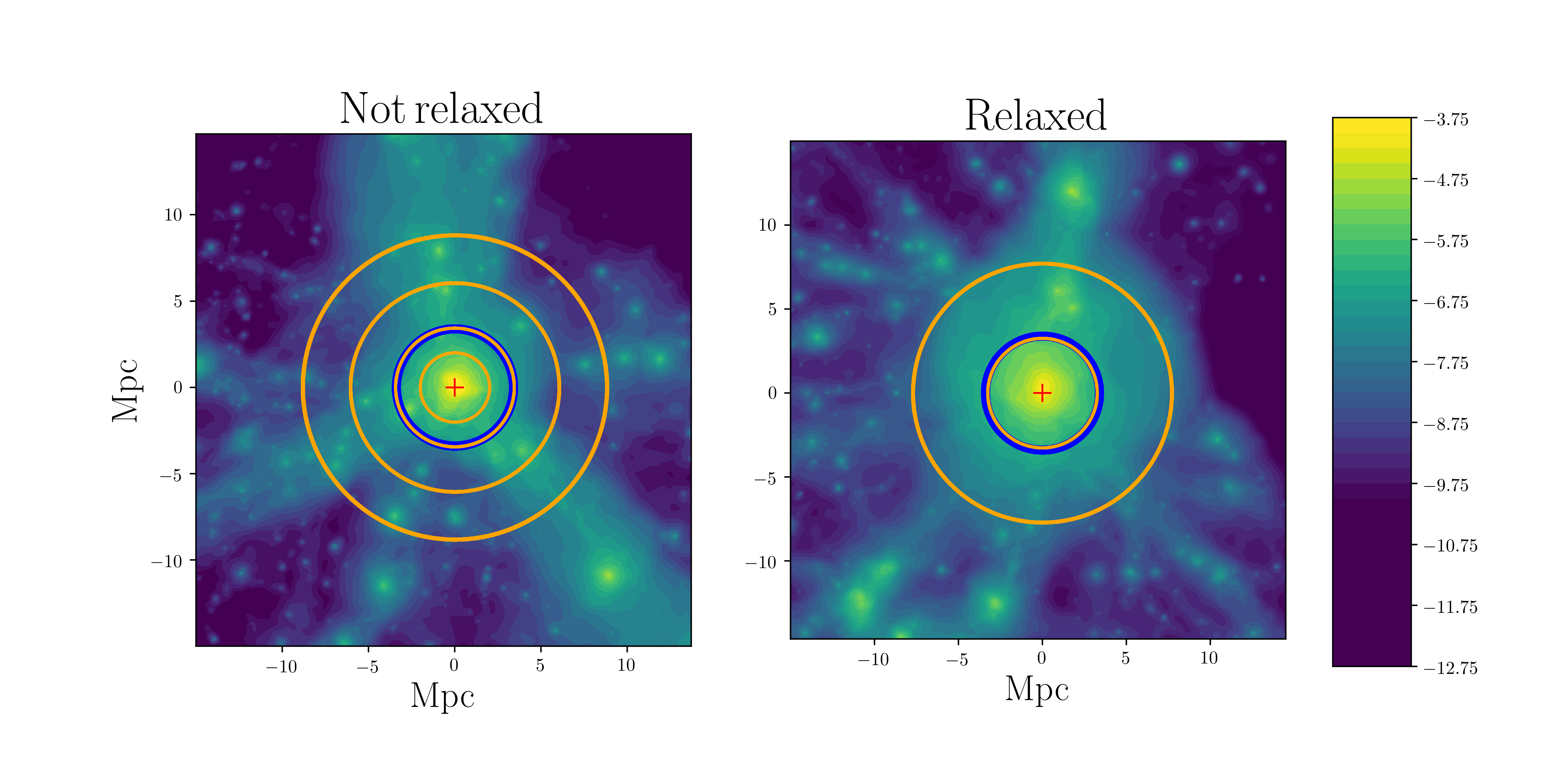}
    \caption{Example Compton $y$ maps from the Three Hundred simulations.  The image on the left shows a cluster with $M_{200m} = 1.45\times10^{15}\,h^{-1}M_{\odot}$ which has been determined to be not relaxed based on the criteria discussed in \S\ref{sec:relaxed}; the figure on the right shows a relaxed cluster with approximately the same mass.  Both clusters are at $z = 0.193$.  The blue and orange rings indicate the locations of $R_{200m}$ and minima in the logarithmic derivative of the cluster $y$ profile, respectively. }
    \label{fig:examples_2d}
\end{figure*}

In several cases, we find it useful to compute the logarithmic derivatives of the measured Compton $y$ profiles.  To do so, we follow the procedure introduced in \citet{Diemer:2014xya}.  For each cluster, we smooth its logarithmic $y$ profile using a Savitsky-Golay filter of window size 5 and polynomial order 2.  The smoothed profile is then fit with a cubic spline, evaluated on a very fine radial grid, and the derivative is computed with finite differences.  We find that our results are not very sensitive to the window size or polynomial order of the Savitsky-Golay filter, provided these are not chosen to introduce so much smoothing that the profile features are washed out.

\begin{figure*}
    \centering
    \includegraphics[scale = 0.5]{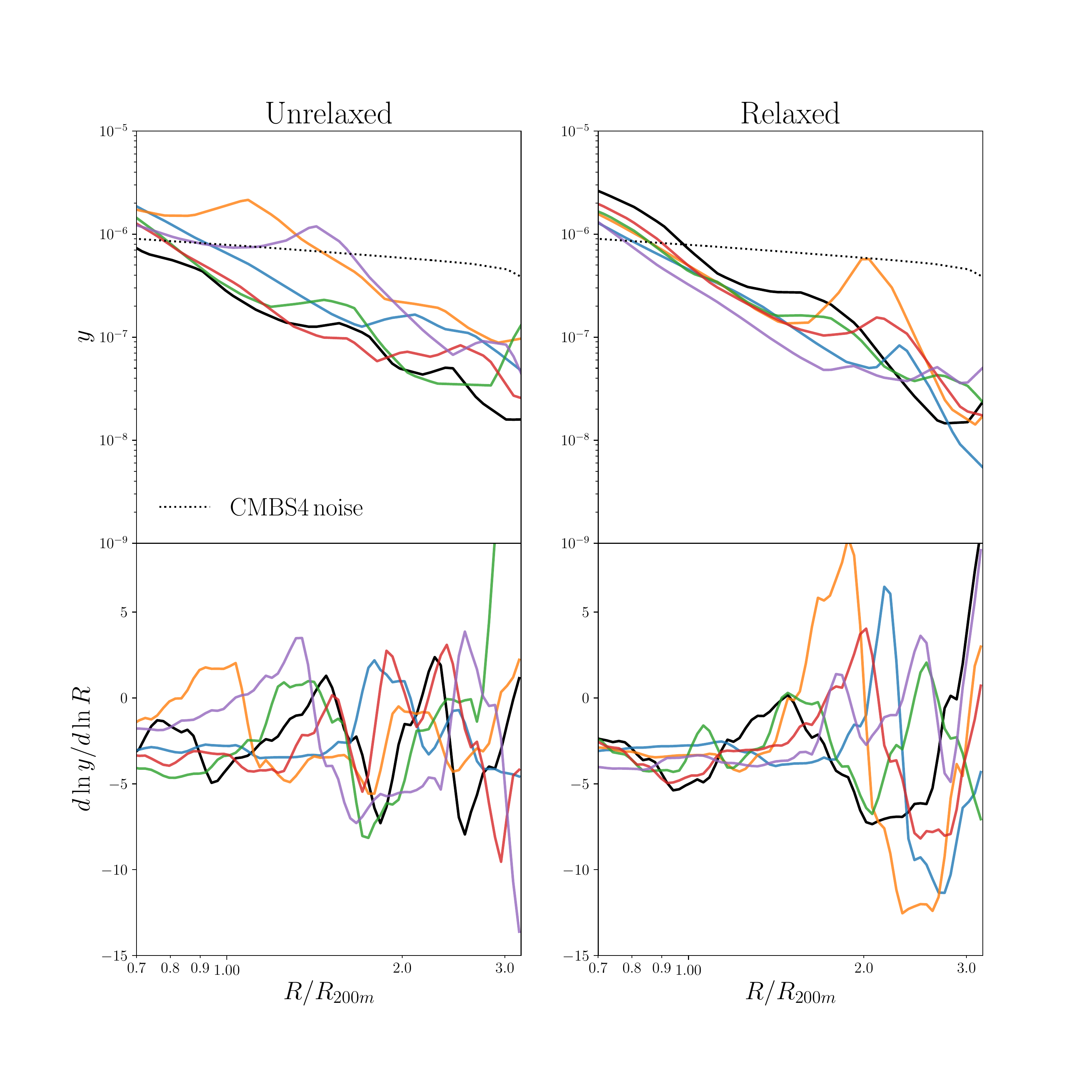}
    \caption{Example Compton $y$ profiles of not relaxed (left) and relaxed (right) clusters.  Top panels shows the Compton $y$ profiles, while the bottom panels show the logarithmic derivatives of this quantity.  Steepening of the $y$ profiles of the relaxed clusters at roughly $2 R_{\rm 200m}$ is evident for many of the clusters. The black lines correspond to the profiles of the two clusters shown in Fig.~\ref{fig:examples_2d}. 
    } 
    \label{fig:individual_profiles}
\end{figure*}

\section{Results}
\label{sec:results}

\subsection{The Compton $y$ profiles of individual clusters}

We first consider the Compton $y$ profiles of individual clusters, arguing that features due to external shocks can be seen in these profiles.  However, we will show that detecting these features in data from future SZ surveys will require averaging across clusters to increase the signal-to-noise, motivating the subsequent analysis discussed in \S\ref{sec:stacking}.  Fig.~\ref{fig:examples_2d} shows example Compton $y$ maps for an not relaxed (left) and relaxed (right) cluster; the determination of relaxation state is described in \S\ref{sec:relaxed}.  The clusters have close to the same mass, $M_{200m} \approx 1.45\times10^{15}\,h^{-1}M_{\odot}$.  The central crosshair indicates the location of the halo center as determined from the maximum density peak in \textsc{AHF}.  The not relaxed cluster has a less circular and less extended $y$ signal, shows an indication of substructures, and exhibits a larger contribution from nearby filaments.  The blue circle in each panel indicates $R_{200m}$ for the clusters.  We will return to the other colored circles below.

Fig.~\ref{fig:individual_profiles} shows a sample of the cluster $y$ profiles (top panels) for some of the not relaxed (left) and relaxed (right) clusters in the full sample.  The black lines indicate the profiles for the two clusters seen in Fig.~\ref{fig:examples_2d}.   We have normalized the radial coordinates relative to the $R_{200m}$ of each cluster, as discussed in \S\ref{sec:profiles}.  There is clearly large scatter between the profiles, as the  clusters have different masses and different levels of nearby structure. The not relaxed clusters typically exhibit more scatter and more fluctuations in their $y$ profiles owing to the presence of large nearby halos.

\begin{figure}
    \centering
    \includegraphics[scale =0.45]{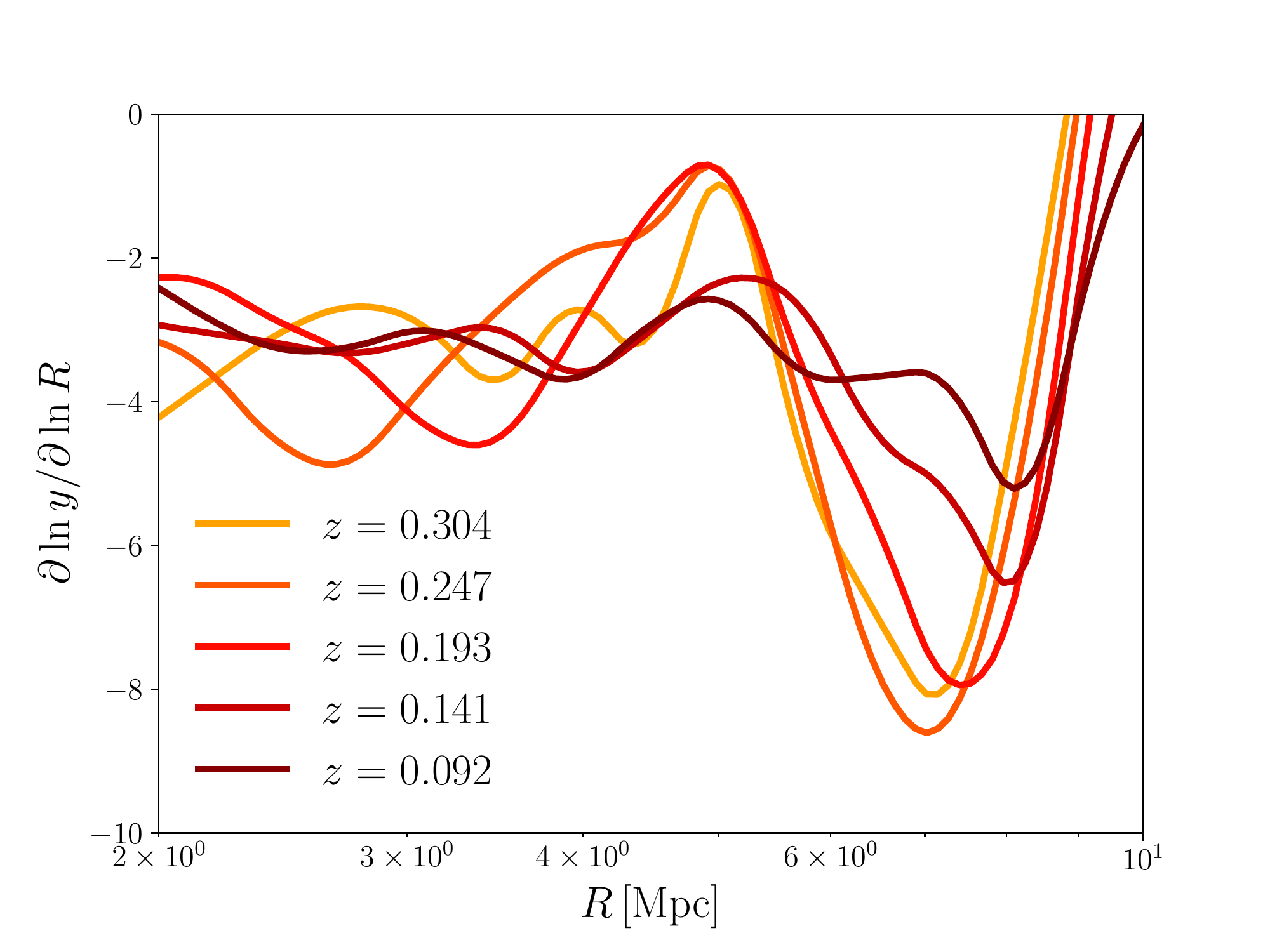}
    \caption{Evolution of the logarithmic derivative of the Compton $y$ profile.  The profiles shown correspond to the relaxed cluster in Fig.~\ref{fig:examples_2d}.
    }
    \label{fig:multiz}
\end{figure}

Gas within the external shock boundary has significantly higher pressure than gas outside the shock.  Consequently, such shocks are associated with a sudden decline in the cluster  pressure profile as one moves away from the cluster center.  This decline is often preceded by a flattening of the pressure profile, as one nears the recently shocked gas \citep{Molnar:2009}.  Since $y$ is proportional to the line-of-sight integral of the pressure profile, we expect shocks to lead to similar behavior in the $y$ profile, albeit smoothed out by the line-of-sight integration.  Indeed, in several of the profiles of the relaxed clusters shown in Fig.~\ref{fig:individual_profiles}, there is an apparent flattening of the profile, followed by a subsequent steepening at roughly $2 R_{200m}$.  This feature is especially clear for the relaxed cluster from Fig.~\ref{fig:examples_2d} (shown with the black curves in the right-hand panels of Fig.~\ref{fig:individual_profiles}), but can be easily seen for the other relaxed clusters, and some of the not relaxed clusters, as well.  These features, and their location at roughly $2R_{200m}$, are consistent with the shock features identified in the analysis of \citet{Molnar:2009}.  There is also a weaker minimum in the logarithmic derivative of several clusters near $R_{200m}$.

In order to more clearly see the flattening and subsequent steepening of the Compton $y$ profile that is associated with shocks, we take the logarithmic derivatives of the $y$ profiles using the procedure described in \S\ref{sec:profiles}; these are shown in the bottom panels of Fig.~\ref{fig:individual_profiles}.  The logarithmic derivative profiles for several clusters exhibit minima near to $2R_{200m}$.  We note that some of the relaxed and not relaxed clusters exhibit minima in their logarithmic derivative profiles at multiple radii. This is not surprising: clusters can have multiple shocks, and some features in the $y$ profiles may not be due to shocks at all, but rather to variations in the gas density around the clusters.  The orange rings in Fig.~\ref{fig:examples_2d} indicate the locations of minima in the logarithmic derivative profiles of the clusters.

Fig.~\ref{fig:individual_profiles} also shows the  noise level projected for $y$ profile measurements with CMB-S4 (dotted black line).  This noise curve represent the standard deviation of the profile from noise alone, where the noise is calculated using Gaussian realization of the $y$ noise power spectrum described in \S\ref{sec:simulated_data}. Note that the noise profile is sensitive to the radial binning: finer bins will have higher noise variance as they average over a smaller area of sky.  For the purposes of this plot, we adopt 10 bins between $0.1$ and $4R_{200m}$.  It is clear from Fig.~\ref{fig:individual_profiles} that it will be difficult to directly measure shock features around individual clusters in data from  CMB-S4: the standard deviation of the noise is larger than the amplitude of the $y$ profile at $2R_{200m}$, where the shocks features occur.  This motivates our decision to focus on the Compton $y$ profiles averaged over many galaxy clusters.  However, detection of shock features in the $y$ profiles of individual clusters may still be possible, particularly for features at lower radial distances (such as virial shocks) and for very massive clusters.  Indeed \citet{Hurier:2017} have recently claimed a detection of a virial shock around a massive cluster using {\it Planck} $y$ maps.

In addition to measuring the cluster $y$ profiles at fixed redshift, it is also useful to consider the evolution of the logarithmic derivative profiles as a function of redshift.  Taking as an example the relaxed cluster shown in the right panel of Fig~\ref{fig:examples_2d},  Fig.~\ref{fig:multiz} shows the evolution of the logarithmic derivative of this cluster's $y$ profile over the range of $z\approx 0.3$ to $z \approx 0.1$.    Although nearby matter (halos and gas) is infalling {\it into} the cluster, the  evolution of the logarithmic derivative minimum is a movement {\it outward}.  This is consistent with the association of the minimum in the logarithmic derivative with cluster shock features, since the shock boundary is expected to grow as the cluster accumulates more gas.

\begin{figure}
    \centering
    \includegraphics[scale = 0.45]{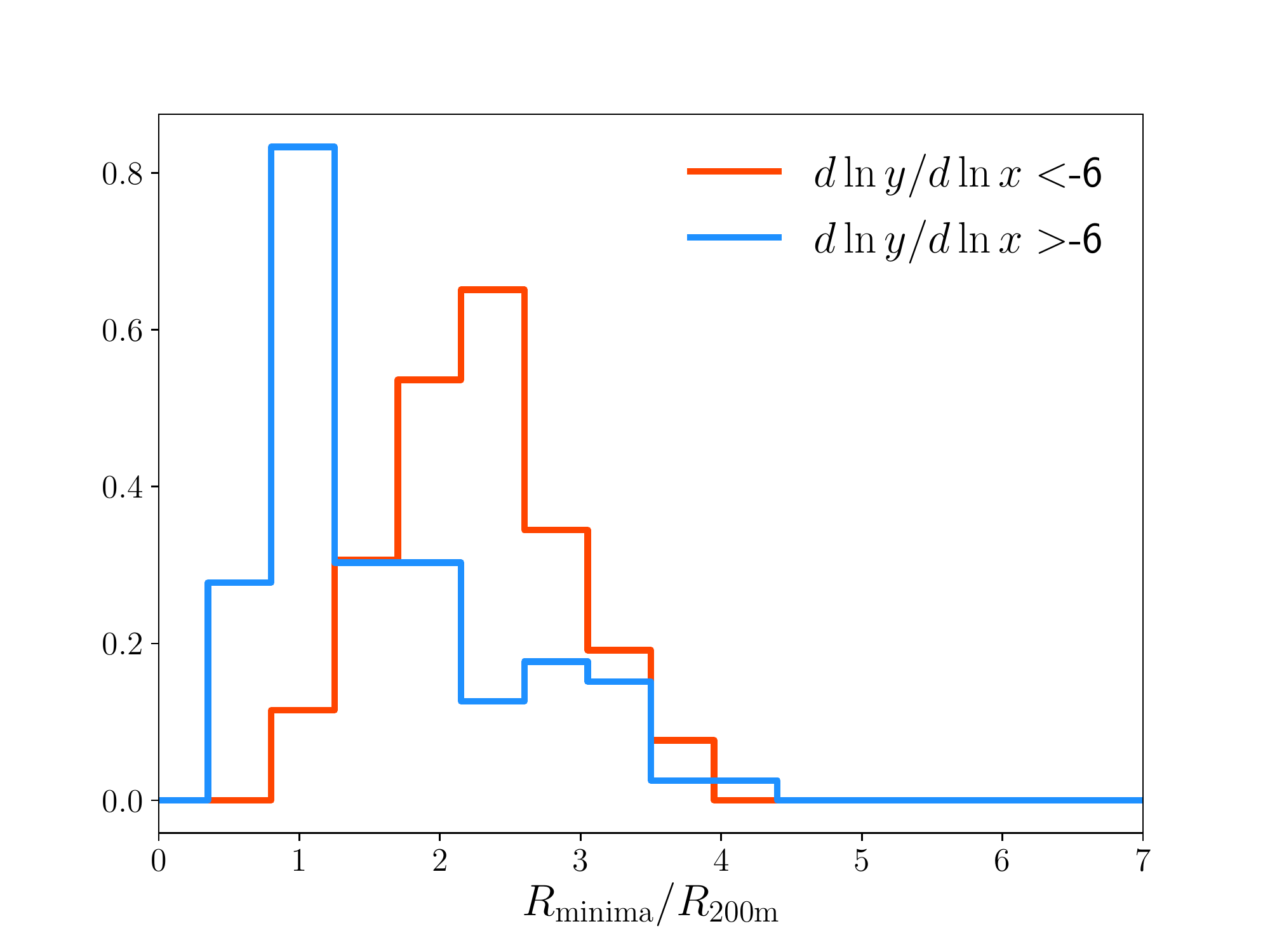}
    \caption{The distribution of locations of minima in the logarithmic derivatives of the Compton $y$ profiles of the simulated clusters.  Minima tend to occur at roughly $R_{200m}$ and $2R_{200m}$.  The minima near $2 R_{200m}$ are typically deeper (corresponding to more negative logarithmic slopes) than those near $R_{200m}$.
    }
    \label{fig:minima_distribution}
\end{figure}

We next consider the distribution of minima in the logarithmic derivatives of the $y$ profiles of all the simulated clusters.  Fig.~\ref{fig:minima_distribution} shows this distribution, split on the depth of the minima in the logarithmic derivative.  Note that a single cluster can have multiple minima that appear in this figure.  The deep minima with slopes less than $-6$ group together around $2R_{200m}$, which we associate with the external shocks.  The clusters also show a preference for shallow minima at roughly $R_{200m}$, which may be due to so-called virial shocks \citep{Molnar:2009}.  

While minima in the logarithmic derivative profiles are (at least sometimes) associated with shocks, we emphasize that the locations of these minima do not necessarily coincide exactly with the locations of shocks identified from the peaks of the entropy profile (e.g., as in  \citealt{Lau:2015}).  For one, $y$ is a projected quantity while the peak of the entropy profile is measured in 3D.  This means that the $y$ profile, unlike the 3D entropy profile, can be significantly impacted by line of sight projections of nearby structure. Furthermore, as seen in \citet{Molnar:2009}, even in 3D, the entropy profile does not peak exactly at the location of the steepest slope of the pressure profile.  
Nonetheless, as we show in \S\ref{sec:other_halo_properties}, it appears that the location of the minimum in the logarithmic derivative of the \textit{cluster-averaged} $y$ profile agrees well (to tens of percent in radius) with the location of the maximum (in absolute value) of the cluster-averaged gas infall velocity, i.e. where the external shock is expected.  This suggests that when averaging over many clusters, the projected radius of the minimum in the logarithmic derivative of $y$ is in rough agreement with the 3D radius of the external shock.

\begin{figure}
    \centering
    \includegraphics[scale = 0.45]{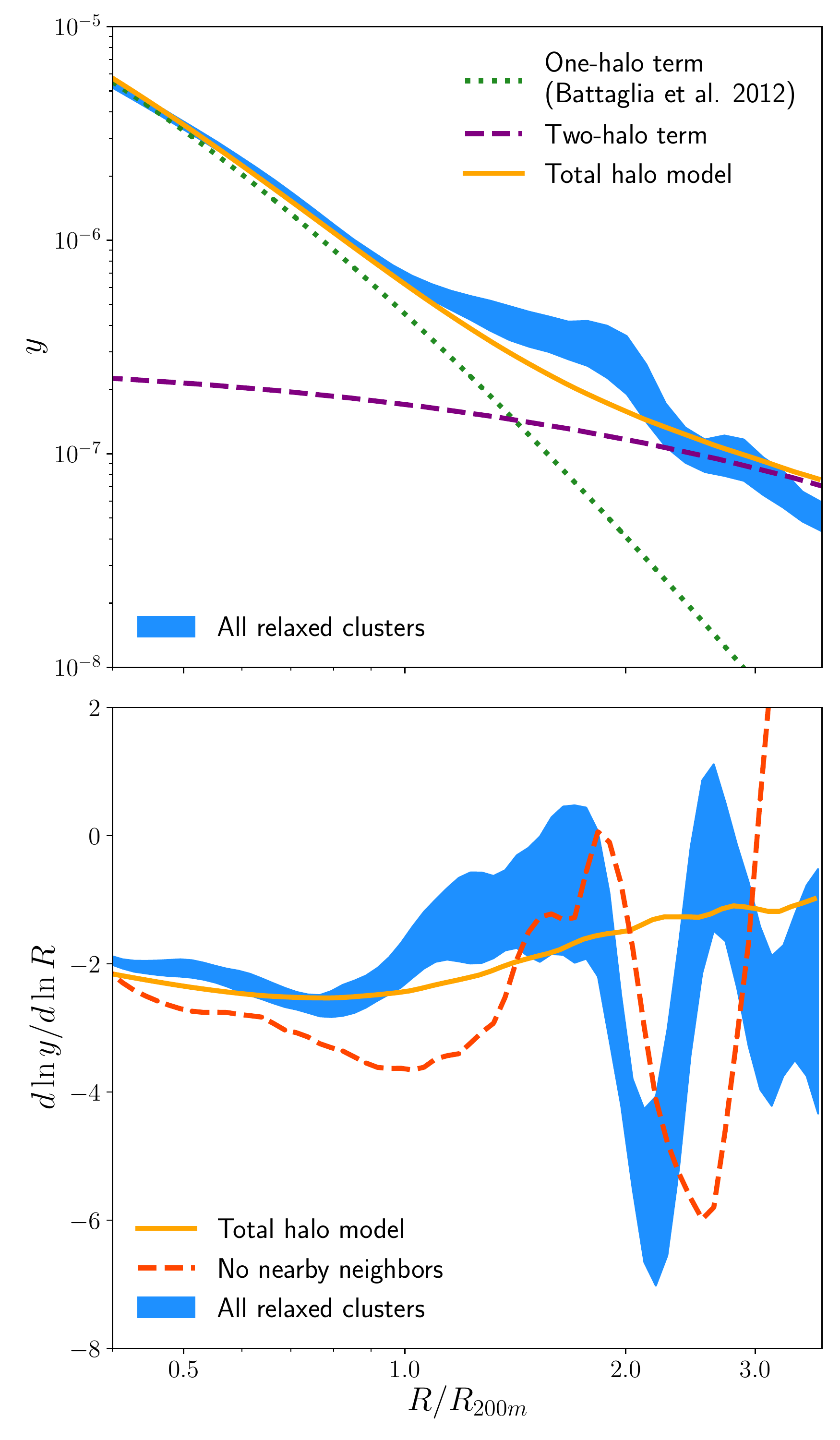}
    \caption{{\it Top}: the averaged Compton $y$ profile of the relaxed clusters as a function of $R/R_{200m}$ (blue band).  The width of the band represents the 68\% confidence interval, computed by jackknifing the cluster sample.  The green dotted curve shows a cluster $y$ profile computed from the \citet{Battaglia:2012} fitting function (i.e. the one-halo term), while the purple dashed curve represents the expected contributions from nearby halos (i.e. the two-halo term).  The orange curve represents the sum of the one and two-halo terms.  The stacked cluster profile exhibits a plateau at $R/R_{200m} \sim 2$, followed by a rapid decline.  We associate this feature with outer shocks in the individual cluster $y$ profiles. {\it Bottom}: logarithmic derivative of the stacked Compton $y$ profile, which shows a narrow minimum at  $R/R_{200m} \sim 2$.  The orange curve represents the logarithmic derivative of the sum of the one and two-halo terms from the top panel.  The red dashed curve shows the  measured logarithmic derivative profile when the cluster sample is restricted to those clusters that do not have massive nearby neighbors.    }
    \label{fig:stacked_profile}
\end{figure}

\subsection{The stacked Compton $y$ profiles}
\label{sec:stacking}

Motivated by the discussion above, we now explore the possibility of averaging $y$ profiles from multiple clusters to improve the signal-to-noise of external shock measurements. It is not obvious that averaging the $y$ profiles of multiple clusters is a sensible way to improve the signal-to-noise of shock measurements.  As seen in Fig.~\ref{fig:individual_profiles}, there is considerable scatter in the $y$ profiles of clusters, and this scatter could cause the shock features to disappear in the stack.  However, we will show below that as long as the profiles are rescaled by $R_{200m}$, a clear feature emerges in the stacked Compton $y$ profiles that is associated with external shocks.

The top panel of  Fig.~\ref{fig:stacked_profile} shows the uncertainty band on the average Compton $y$ profile of the relaxed sample of clusters, computed as described in \S\ref{sec:profiles}. The profile exhibits a flattening, followed by a rapid steepening at roughly $2R_{200m}$.   The bottom panel of Fig.~\ref{fig:stacked_profile} shows the constraints on the averaged logarithmic derivative profile.  There is a deep minimum in the logarithmic derivative profile near $2R_{200m}$, which we associate with external shocks.   The appearance of this feature in the stacked Compton $y$ profile is one of the main results of our analysis.  We note that the feature still appears to some extent when stacking the profiles of not relaxed clusters, but it is less pronounced.  When stacking the cluster profiles without rescaling by $R_{200m}$, the sharp minimum feature again becomes weaker; again, this reflects the greater universality of the outer cluster profiles when scaled by $R_{200m}$ \citep{Lau:2015}.  We note that the shallow minimum seen at roughly $R_{200m}$ may be due to the inner (virial) shock.  

The green dotted curve in the top panel of Fig.~\ref{fig:stacked_profile} shows the expected $y$ profile derived from the \citet{Battaglia:2012} pressure profile.  The \citet{Battaglia:2012} profile was derived from fitting the inner part of the cluster pressure profiles in hydrodynamical simulations, so it is not surprising that it agrees with our results very well at small scales. At large scales, however, the profile departs from the measurements as a result of the shock feature, and the contributions from nearby halos, which we discuss in more detail below.

\subsection{The impact of nearby halos on the $y$ profile}
\label{sec:nearbyhalos}

One might worry that the presence of nearby halos could impact our ability to detect features associated with external shocks in the cluster-averaged Compton $y$ profile.  As shown in \citet{Power2020}, massive infalling substructures can produce clear signatures of shocks in the direction of their motions.  Furthermore, overdensities of gas in nearby halos could produce features in the $y$ profile that appear similar to the features caused by shocks.  We will argue below that while nearby halos do have an impact on the Compton $y$ profiles of clusters, they do not significantly impact our main conclusions about the shock features in the average $y$ profile of many clusters.

In the language of the halo model \citep{Cooray:2002}, the contributions to the cluster-centric $y$ profile coming from nearby halos is the two-halo term.  The two-halo term in the context of the halo-$y$ correlation has been considered by several authors \citep[e.g.][]{Komatsu:2002,Vikram:2017}.   The expected two-halo term for the clusters in the sample is shown as the purple dashed curve in the top panel of Fig.~\ref{fig:stacked_profile}.  We compute the two-halo term as described in \citet{Pandey:2019}, adopting the \citet{Battaglia:2012} pressure profile model and a linear bias model to describe halo clustering.  The two-halo term offers a reasonable description of the $y$ profile at large $R/R_{200m}$, despite simplifications such as our  assumption of linear bias.   The total halo model profile is the sum of the one and two-halo terms, shown as the orange curve in the top and bottom panels of Fig.~\ref{fig:stacked_profile}.  While the shock feature in the cluster-averaged $y$ profile occurs near to the one-to-two halo transition, it appears qualitatively different from the halo model prediction for this transition.  The shock feature results in a plateauing of the $y$ profile, followed by a steep decline, followed by the resumption of a less steep profile (top panel of Fig.~\ref{fig:stacked_profile}).  The  one-to-two halo transition, however, does not exhibit any such steep decline.  The bottom panel of Fig.~\ref{fig:stacked_profile} shows that the logarithmic derivative of the total halo model prediction never goes below -3, while the logarithmic derivative of the simulated cluster profiles reaches slopes as steep as -5 and steeper.  In other words, assuming the cluster pressure profile of \citet{Battaglia:2012}, the halo model fails to provide an accurate description of shock features in the Compton $y$ profile.  This is not surprising, as the halo model prediction includes no information about the shocking of gas.

To test the impact of large nearby halos on the inferred shock locations, we repeat our analysis after removing any of the clusters that have a neighbor with $M > 8\times 10^{13} M_{\odot}$ within 10 Mpc.  This selection reduces the number of relaxed clusters to 13.  A comparison between our results with all relaxed clusters and the results found when additionally excluding clusters with massive neighbors is shown in the bottom panel of Fig.~\ref{fig:stacked_profile}.  Since the exclusion criterion significantly changes the cluster sample, we do not expect the $y$ profile to remain unchanged.  Indeed, selecting halos without nearby massive neighbors amounts to a selection on cluster environment, which correlates strongly with accretion rate.  We expect then, that this selection may modify the shock location to some extent (see further discussion of this point in \S\ref{sec:splashback}).  Nonetheless, Fig.~\ref{fig:stacked_profile} shows that the appearance of the shock feature is fairly robust to the selection of clusters without massive neighbors.  In principle, one could also attempt to remove two-halo contributions to the cluster-$y$ profile by selecting isolated clusters; however, see \citet{Hill:2018} for an explanation of why this approach presents additional difficulties.

\begin{figure}
    \centering
    \includegraphics[scale = 0.44]{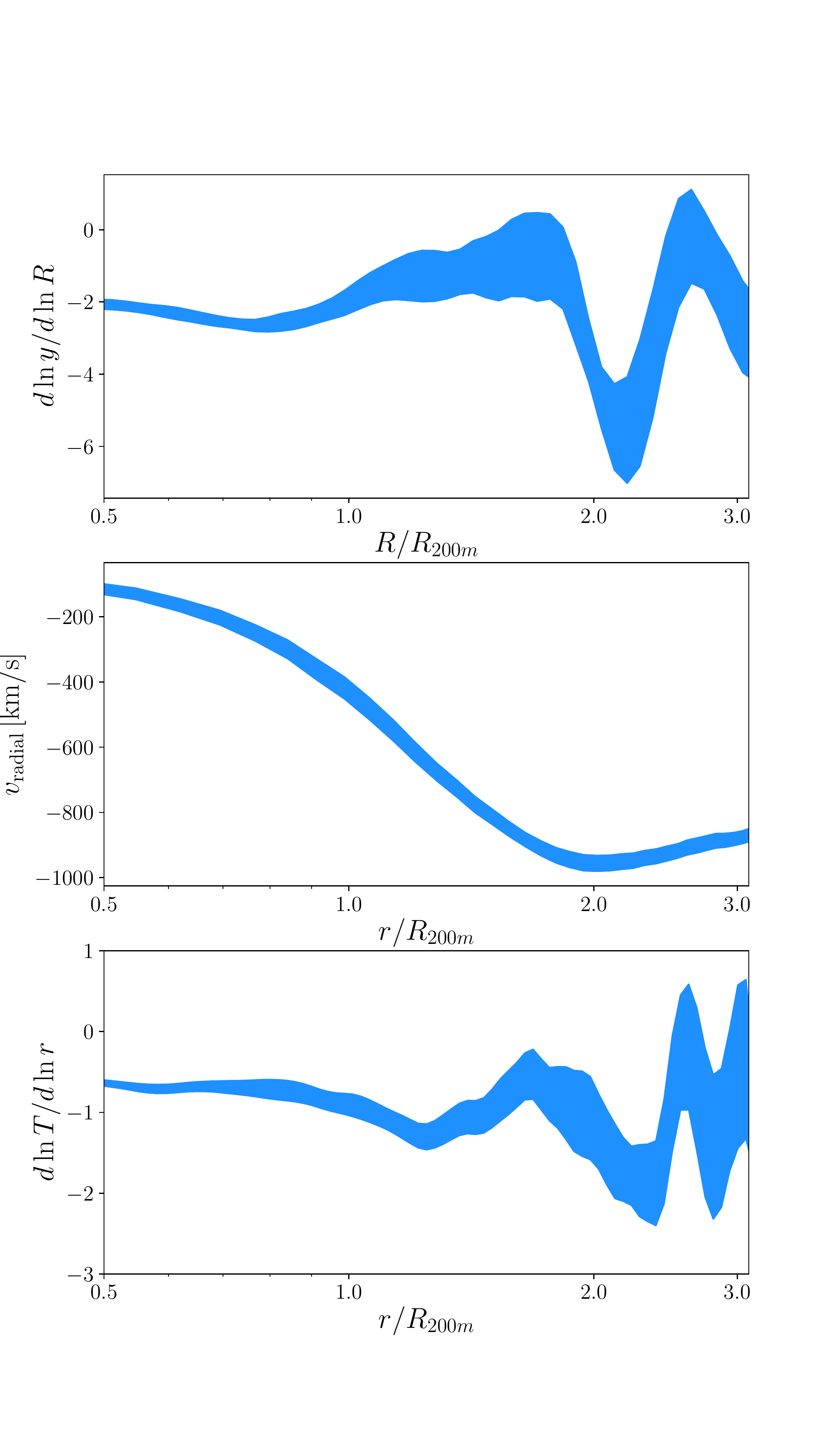}
    \caption{{\it Top}: logarithmic derivative of the average Compton $y$ profile of the relaxed clusters.  The profile is plotted as a function of the projected radius, $R/R_{200m}$.  {\it Middle}: the averaged radial velocity profile of gas around the same sample of clusters.  The location of the narrow minimum in the logarithmic derivative of Compton $y$ (seen in the top panel) roughly aligns with the minimum in the radial velocity profile, as expected because of shocks experienced by the infalling gas. The profile is plotted as a function of the 3D cluster-centric radius, $r/R_{200m}$.  {\it Bottom:} the logarithmic derivative of the averaged radial temperature profile of gas around the same sample of clusters.  The temperature profiles exhibit a steepening near the shock radius.  }
    \label{fig:stacked_profile_velocity_temp}
\end{figure}

\subsection{Connecting the $y$ profile with other halo properties}
\label{sec:other_halo_properties}

Accreting gas rapidly decelerates and increases in temperature as it passes through the external shock.  Consequently, we expect the shock features observed in the $y$ profile to be related to features in the gas velocity and temperature profiles.  Fig.~\ref{fig:stacked_profile_velocity_temp} explores these connections.  

The top panel of Fig.~\ref{fig:stacked_profile_velocity_temp} shows the logarithmic derivative of the cluster-averaged $y$ profile, as in Fig.~\ref{fig:stacked_profile}.  For this figure, we consider only the relaxed sample of clusters.  The middle panel of Fig.~\ref{fig:stacked_profile_velocity_temp} shows the average radial velocity profile of the gas for the same set of clusters.  The narrow minimum in logarithmic derivative of Compton $y$ closely corresponds with the minimum in the radial velocity profile.  This supports our interpretation of the feature in the averaged $y$ profile as being due to the outer shock. Note we use two different radii in  Fig.~\ref{fig:stacked_profile_velocity_temp}: the $y$ profile is a projected quantity and we therefore plot it as a function of the projected radial distance, $R$, while the radial velocity profile and temperature profiles are measured as functions of the three-dimensional radial distance, $r$.

The bottom panel of Fig.~\ref{fig:stacked_profile_velocity_temp} shows the logarithmic derivative of the averaged cluster temperature profile.  As one moves to larger radius near the shock location, the temperature profile becomes steeper and steeper.  This is expected, as the temperature of the gas increases as it passes through the shock.

\begin{figure}
    \centering
    \includegraphics[scale = 0.45]{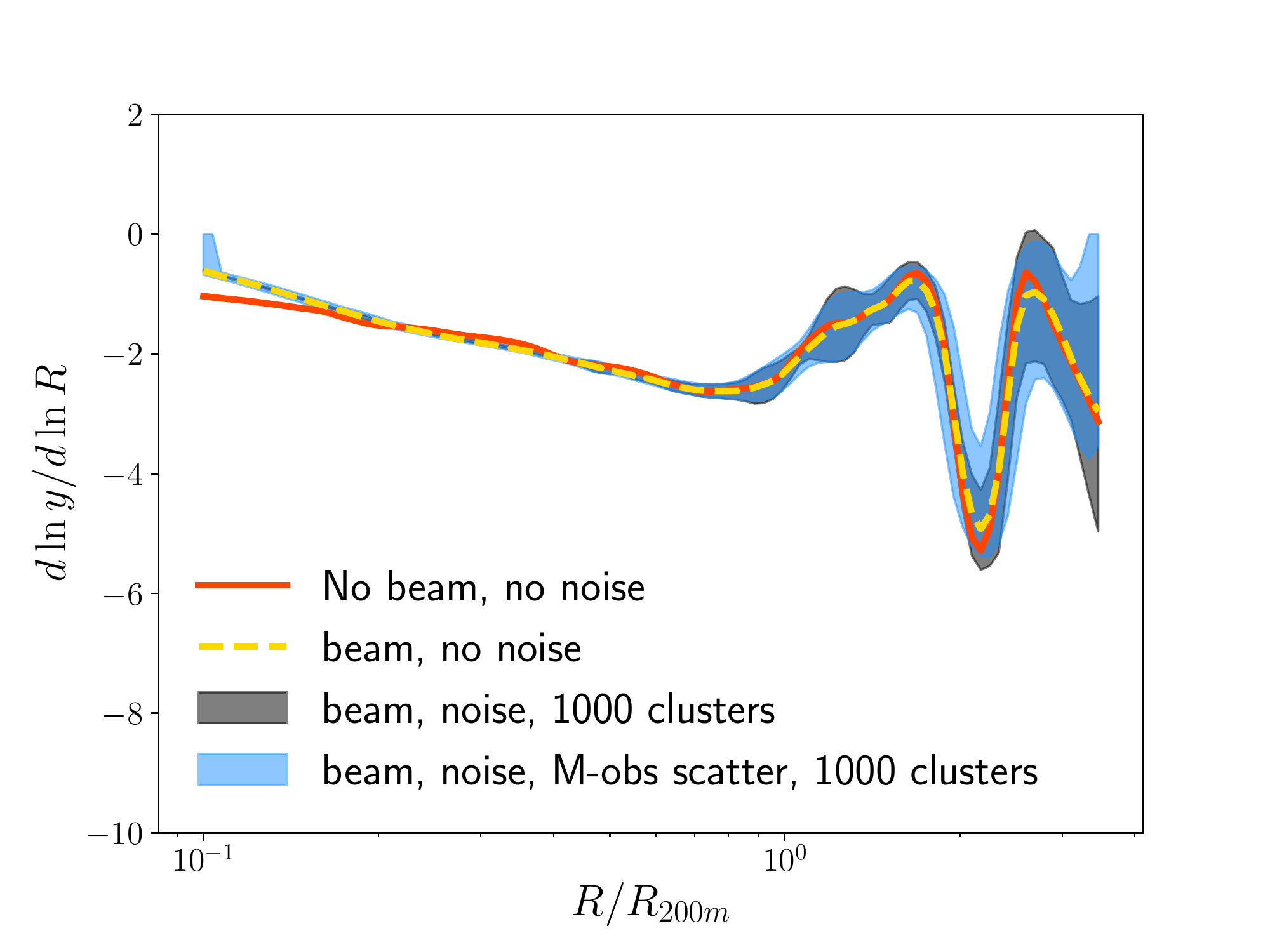}
    \caption{The impact of telescope beam and observational uncertainties on measurements of the logarithmic derivative of the cluster-averaged Compton $y$ profile.  The red curve shows the true log-derivative of the average Compton $y$ profile of the simulated galaxy clusters.  The orange dashed curve shows the impact of smoothing of the Compton $y$ maps by a 1 arcmin beam.  The black band shows the expected errorbars resulting from CMB-S4-like noise when averaging across 1000 clusters (note that the band does not include scatter due to variations in the profile shapes).  The blue band additionally adds scatter in the mass-observable relation.
    }
    \label{fig:mock_obs}
\end{figure}

\subsection{Impact of observational uncertainties}

We now consider how three observational effects can impact our results: (1) the instrumental beam, (2) noise in the Compton $y$ maps, (3) uncertainty on the value of $R_{200m}$ used when rescaling the profiles.   Fig.~\ref{fig:mock_obs} shows the impact of these observational effects.  The red curve shows the stacked logarithmic derivative profile in the absence of beam or noise.  The gold curve adds the impact of the instrumental beam.  The beam flattens the profile in the inner parts, leading to a more positive logarithmic derivative.  Even with the beam, though, the narrow minimum in the logarithmic derivative of the Compton $y$ profile is clearly visible.   This is because for our clusters at $z = 0.193$, the shock extends a sufficiently large angle on the sky that it is not significantly degraded by the beam.

\begin{figure*}
    \centering
    \includegraphics[scale=0.5]{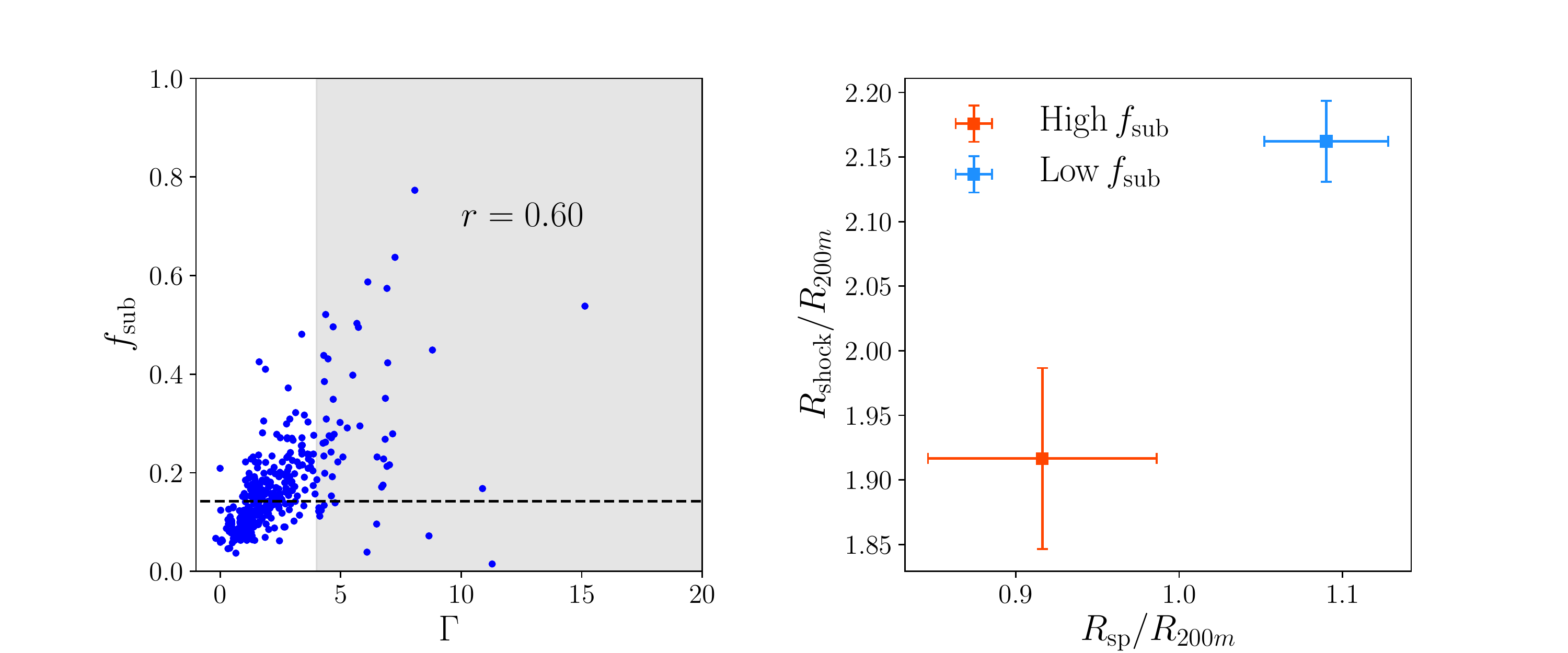}
    \caption{
    {\it Left}: the quantity $f_{\rm sub}$ that we use to identify the relaxed population of clusters is correlated with the cluster accretion rate, $\Gamma$, with correlation coefficient of $r=0.6$.  The grey band illustrates our cut on cluster accretion rate imposed when computing the measurements shown in the right panel.  The dashed horizontal line indicates the median $f_{\rm sub}$ of the remaining clusters, which is used to split the full sample into the high and low $f_{\rm sub}$ subsamples shown at right.  {\it Right}:  the shock radius and splashback radius are correlated.  The shock radius here is identified using the minimum of the logarithimic derivative of the cluster-averaged $y$ profile, while the splashback radius is identified from the dark matter particles as described in the main text.  The two data points shown correspond to the cluster samples defined in the left panel. } 
    \label{fig:rsp_vs_rshock}
\end{figure*}

The black band shown in Fig.~\ref{fig:mock_obs} represents the uncertainty on the logarithmic derivative that results from using a sample of 1000 clusters when instrumental noise is included.  We find that 1000 clusters are sufficient to obtain a reliable reconstruction of the Compton $y$ profile, assuming noise levels characteristic of CMB-S4.  Since current generation surveys like SPT-3G \citep{Benson:2014} and AdvACT \citep{Henderson:2016} are already detecting thousands of clusters, and since these surveys have (in some cases) noise levels that are only factors of a few larger than CMB-S4, these surveys may already be well positioned to detect a signature of external shocks through stacking.  Our analysis here focuses on CMB-S4 in order to bracket the possibilities of current and future surveys. 

The blue band in Fig.~\ref{fig:mock_obs} incorporates additional uncertainty due to scatter in the relationship between cluster observable and cluster mass, as described in \S\ref{sec:mock_obs}.  This additional scatter smears out the stacked $y$-profile, since the profiles are stacked in terms of an assumed $R_{200m}$, which is in turn determined from the mass-observable relation.   The smearing of the $y$ profile results in a weakening of the sharp changes in logarithmic derivative profile, as seen in Fig.~\ref{fig:mock_obs}.  However, the shock-induced feature can still be detected at high significance.

\subsection{Splashback and shocks: correlation between the splashback radius and shock location}
\label{sec:splashback}

Recently, the splashback radius has been established as the boundary of the virialized, or multi-streaming region of the dark matter component of a halo \citep{Diemer:2014xya,AD14}. This surface traces the boundary of the first apocenter of the most recently accreted material, separating a region of dark matter infall from the regions within which dark matter particles or other collisionless objects like galaxies orbit.  Arguably, the external shock forms the boundary of the gaseous component of the halo. 

It is known that both the locations of external shocks and the locations of the outer caustic in the dark matter distribution both depend on the evolution history of the halo \citep{Bertschinger:1985, Shi:2016}. In particular, both these features are a function of the mass accretion rate of the halo. We therefore expect the splashback radius and the shock location to be correlated.  Although analytical studies with self-similar models \citep{Bertschinger:1985} showed that the location of the accretion shock coincides with the location of the outer caustic, studies in simulations have shown that external shocks can exist far outside the virial boundary of the halo \citep{Molnar:2009,Zhang:2019kej, Zhang:2020}.  In this section we investigate how the shock features in the averaged Compton $y$ profiles correlate with the splashback feature in the dark matter profiles of clusters. 

While the actual splashback surface is ideally measured from the apocenter of particle orbits as in \citet{Diemer:2017ecy}, we measure the splashback radius here as the location of the minimum of the slope in the logarithmic density profile. This radius is known to capture the phase space transition between the infall and virialized region and is moreover accessible observationally \citep[e.g.][]{More:2015ufa}.

We compare the shock and splashback radii for clusters as a function of the accretion history of the halo. We use $f_{\rm sub}$, the mass fraction of the cluster in substructure, as a proxy for the accretion rate. This quantity is potentially observable and is related to the number of bright satellites in clusters. Young halos that are fast accreters have large satellite fractions. With time, as the halo stops growing and enters a slow accretion phase, substructure gets destroyed due to tidal effects and dynamical friction. $f_{\rm sub}$ is therefore strongly correlated to the accretion rate of the cluster as shown in the left panel of Fig. \ref{fig:rsp_vs_rshock}. The accretion rate of the cluster, $\Gamma=d\log M_{200m}/d\log a$, is obtained from the mass accretion history measured from the simulation. We evaluate the slope of the mass history at $z=0.193$ by fitting the function $M=e^{-\alpha z}$ \citep{2002ApJ...568...52W} to the mass between between the $0.193 < z < 0.350$. 

In order to explore the relation between shock and splashback radius, we divide the cluster sample into two subsets with different $f_{\rm sub}$.  We first remove clusters with $\Gamma > 4$ (the grey region of the left panel of Fig.~\ref{fig:rsp_vs_rshock}), as clusters with large $\Gamma$ may be in the process of major mergers.  We then split the remaining cluster sample in halves based on the median value of $f_{\rm sub}$.  The value of $f_{\rm sub}$ used for this split is shown as the horizontal dashed line in the left panel of Fig.~\ref{fig:rsp_vs_rshock}.  

In the right panel of Fig.~\ref{fig:rsp_vs_rshock} we show the relation between the location of the shock and the location of the splashback radius for the two cluster subsamples split on $f_{\rm sub}$.  The shock radius is determined from the location of the minimum of the logarithmic derivative of the averaged $y$ profile, while the splashback radius is determined from the averaged subhalo density profiles.  Both radii are normalized by $R_{200m}$, so any correlation between the cluster mass and the two radii is effectively cancelled. We see that halos with high $f_{\rm sub}$, corresponding to high accretion rate, have both a smaller splashback radius and a smaller shock radius.  The dependence of the splashback radius on accretion rate is expected, as particle apocenters become smaller due to the rapid growth of the halo potential \citep{Diemer:2014xya}.  We see a similar behavior in the location of the outer shock.  The correlations between splashback and shock radius has previously been seen in simulations \citep{Walker:2019, Aung:2020}.  Our result here shows that this correlation persists when the shock location is estimated from the minimum of the logarithmic derivative of the stacked $y$ profile.

Fig.~\ref{fig:rsp_vs_rshock} also makes it clear that the inferred shock location will vary depending on the $f_{\rm sub}$ cuts.  Our nominal cut of $f_{\rm sub} < 0.1$ corresponds (very) roughly to the blue point in the right panel of Fig.~\ref{fig:rsp_vs_rshock}.  As we consider clusters with higher $f_{\rm sub}$, the shock location decreases, and the profiles exhibit more scatter, as seen in Fig.~\ref{fig:individual_profiles}.  As a result the errorbars on the shock position for the high $f_{\rm sub}$ sample are larger than those for the low $f_{\rm sub}$ sample.  An analysis in data would need to consider the signal-to-noise trade off between increasing the number of clusters and increasing the scatter in individual cluster $y$ profiles as the $f_{\rm sub}$ (or other accretion rate proxy) cut is relaxed. 

\section{Discussion}
\label{sec:discussion}

\subsection{Summary}

The outskirts of galaxy clusters reflect their recent accretion history and their local environments.  The collisional gas accreting onto clusters experiences shocks at several $R_{200m}$ that significantly impact the thermodynamic properties of the ICM.   We have presented an exploration of the detectability of shock features in the stacked Compton $y$ profiles of massive galaxy clusters.  Because dark matter is collisionless, on the other hand, its accretion gives rise to the splashback boundary of the cluster, and a corresponding feature in the cluster mass profile.  We have explored the correlation between shock features in the cluster $y$ profiles and the splashback features in the cluster mass profiles.  

Gas shocks in cluster outskirts have proven difficult to constrain observationally because of the low gas densities in the cluster outskirts.  One promising route to detection is via the SZ effect.  However, current and future wide-field SZ surveys are unlikely to have the sensitivity to detect these features around individual clusters (except perhaps the most massive ones), as we have shown.
Motivated by these considerations, we show that stacking cluster $y$ profiles as a function of $R/R_{200m}$ can provide a promising route to constraining shock features around future samples of galaxy clusters.  We demonstrate that outer shocks result in a narrow minimum in the logarithmic derivative of the stacked cluster $y$ profile.  This shock-associated feature is robust to smoothing by instrumental beams and uncertainty in $R_{200m}$.  Furthermore, for realistic levels of noise (i.e. typical of CMB-S4), high significance detections of the shock features can be obtained with samples of roughly a thousand clusters.  Such detections may already be in reach of ongoing CMB surveys such as SPT-3G and AdvACT.

\subsection{Implications}

We have shown that shock features are detectable in the averaged $y$ profiles of galaxy clusters with near-term CMB data.  Obtaining such a detection is an exciting prospect.  The locations and properties of shocks are connected to the cluster accretion rate and the cluster's large-scale environment.  Measurement of the shock radius in the averaged $y$ profile could be used to test different proxies for the accretion rate, and to compare to simulations.  Detection of shocks using SZ observations in conjunction with measurements of the cluster splashback radius using weak lensing \citep[e.g.][]{Chang:2018} will provide insight into cluster outskirts in gas and total mass.  Detection of cluster external shocks in large cluster samples may also enable improved constraints on models of cosmic ray acceleration in the shock fronts.  Moreover, the statistical detection of shocks explored here can open up a new window into studying how the properties of galaxies change as they cross the shock and splashback boundaries of clusters \citep[e.g.][]{Baxter:2017, Shin:2019, adhikari2020probing}.

However, the detection of shock features also poses some challenges.  In particular, if these features can be detected at high significance, modelling these features may be necessary for analyzing high-precision observables derived from the SZ effect.  In particular, current analyses of the SZ power spectrum and cross-correlations with the SZ typically do not include shock features when modeling these observables.  Working in physical coordinates (rather than coordinates rescaled by $R_{200m}$) will smear out the shock features, but some impact from the shocks may remain.  We postpone a more careful investigation of the effects of shocks on other observables to future work.

\section*{Acknowledgements}

We thank Dhayaa Anbajagane, Neal Dalal, Andrey Kravtsov, Shivam Pandey, Debora Sijacki and Congyao Zhang for useful discussions related to this paper. WC acknowledges support from the European Research Council under grant number 670193 and further acknowledges the science research grants from the China Manned Space Project with NO. CMS-CSST-2021-A01 and CMS-CSST-2021-B01. AK is supported by the Ministerio de Ciencia, Innovaci\'{o}n y Universidades (MICIU/FEDER) under research grant PGC2018-094975-C21 and further thanks Cocteau Twins for victorialand.

\section*{Data availability}

The data used to generate the figures in this work are available upon request.

\bibliographystyle{mnras}
\bibliography{ref}

\bsp	
\label{lastpage}
\end{document}